\documentclass[twocolumn,showkeys,showpacs, preprintnumbers,amsmath,amssymb]{revtex4}
\usepackage{graphicx}
\usepackage{dcolumn}
\usepackage{bm}
\begin{document}
\title{The effect of core polarization on longitudinal form factors in $^{10}$B}
\author{Fouad~A.~Majeed}\email{fouadalajeeli@yahoo.com}
\affiliation{Department of Physics, College of Education for Pure Science, University of Babylon, P.O.Box 4., Hilla-Babylon, Iraq.}
\date{\today}
\begin{abstract}
Electron scattering Coulomb form factors for the single-particle quadrupole transitions in $p$-shell $^{10}$B nucleus have been studied. Core polarization
effects are included through a microscopic theory that includes excitations from the core orbits up to higher orbits with 2$\hbar$$\omega$ excitations. The
modified surface delta interaction (MSDI) is employed as a residual interaction. The effect of core polarization is found essential in both the transition
strengths and momentum transfer dependence of form factors, and gives a remarkably good agreement with the measured data with no adjustable parameters.
\end{abstract}
 \maketitle
\section{Introduction}
Comparisons between calculated and measured longitudinal electron scattering form factors have long been used as stringent tests of models of nuclear
structure~\cite{FA06,FA07}. Shell model within a restricted model space succeeded in describing static properties of nuclei, when effective charges are
used. The Coulomb form factors have been discussed for the stable sd-shell nuclei using sd-shell wave functions with phenomenological effective
charges~\cite{RA83}. For p-shell nuclei, Cohen–Kurath~\cite{CK65} model explains well the low-energy properties of p-shell nuclei. However, at
higher-momentum transfer, it fails to describe the form factors. Radhi {\em et al.}~\cite{RA01,RA02,RA03,R03,RJ03} have successfully proved that the
inclusion of core polarization effects in the $p$-shell and $sd$-shell are very essential to improve the calculations of the form factors. Restricted
1$p$-shell models were found to provide good predictions for the $^{10}$B natural parity level spectrum and transverse form factors~\cite{CD95}. However,
they were less successful for $C2$ form factors and give just 45\% of the total observed $C2$ transition strength. Expanding the shell-model space to
include 2$\hbar$$\omega$ configurations in describing the form factors of $^{10}$B, Cichocki {\em et al.}~\cite{CD95} have found that only a 10\%
improvement was realized. The purpose of the present work is to study the $C2$ form factors for $^{10}$B by including higher-energy configurations as a
first-order core polarization through a microscopic theory which combines shell model wave functions and highly excited states. Single-particle wave
functions are used as a zero-th contribution and the effect of core polarization is included as a first-order perturbation theory with the modified surface
delta interaction (MSDI)~\cite{PM77} as a residual interaction and a 2$\hbar$$\omega$ for the energy denominator. The single-particle wave functions are
those of the harmonic-oscillator (HO) potential with size parameter \emph{b} chosen to reproduce the measured root mean square (rms) charge radii of these
nuclei.
\section{Theory}
The core polarization effect on the form factors is based on a microscopic theory, which combines shell model wave functions and configurations with higher
energy as first order perturbations; these are called core polarization effects. The reduced matrix elements of the electron scattering operator
$T$${_\Lambda}$ is expressed as the sum of the product of the elements of the one-body density matrix (OBDM)
$\chi^{\Lambda}_{\Gamma_{f}\Gamma_{i}}$($\alpha$, $\beta$) times the single-particle matrix elements, and is given by
\begin{equation}
\langle{\Gamma}_{f}|||T_{\Lambda}|||{\Gamma}_{i}\rangle=\sum_{\alpha,\beta} \chi^{\Lambda}_{\Gamma_{f}\Gamma_{i}}(\alpha,
\beta)(\alpha|||T_{\Lambda}|||\beta),
\end{equation}
where $\alpha$ and $\beta$ label single-particle states (isospin is included) for the model space. For p-shell nuclei, the orbits $1p_{3/2}$ and $1p_{1/2}$
define the model space. The states $\mid$${\Gamma}_{i}$$\rangle$ and ${\Gamma}_{f}$ are described by the model space wave functions. Greek symbols are used
to denote quantum numbers in coordinate space and isospace, i.e. ${\Gamma}$$_{i}$$\equiv$$J_{i}$$T_{i}$,~${\Gamma}$$_{f}$$\equiv$$J_{f}$$T_{f}$ and
$\Lambda$=$JT$. According to the first-order perturbation theory, the single-particle matrix element is given by~\cite{PM77}
\begin{eqnarray}
(\alpha|||T_{\Lambda}|||\beta) &=&\langle\alpha|||T_{\Lambda}|||\beta\rangle
+\langle\alpha|||T_{\Lambda}\frac{Q}{E_{i}-H_{0}}V_{res}|||\beta\rangle\nonumber\\
&+&\langle\alpha|||V_{res}\frac{Q}{E_{f}-H_{0}}T_{\Lambda}|||\beta\rangle.
\end{eqnarray}
The first term is the zeroth-order contribution. The second and third terms are the core polarization contributions. The operator Q is the projection
operator onto the space outside the model space. For the residual interaction, $V_{res}$, we adopt the MSDI~\cite{PM77}. $E_{i}$ and $E_{f}$ are the
energies of the initial and final states, respectively. The core polarization terms are written as~\cite{PM77}
\begin{eqnarray}
\sum_{\alpha_{1},\alpha_{2},\Gamma}\frac{(-1)^{\beta+\alpha_{2}+\Gamma}}{e_{\beta}-e_{\alpha}-e_{\alpha_{1}}+e_{\alpha_{2}}} (2\Gamma+1) \left\{
\begin{array}{lll}
\alpha & \beta & \Lambda\\
\alpha_{2}& \alpha_{1}& \Gamma\end{array}\right\}\nonumber\\
\times\sqrt{(1+\delta_{\alpha_{1}\alpha})(1+\delta_{\alpha_{2}\beta})}\langle\alpha\alpha_{1}|V_{res}|\beta\alpha_{2}\rangle\langle\alpha_{2}|||T_{\Lambda}|||\alpha_{1}\rangle\nonumber\\
+\texttt{terms with $\alpha_{1}$ and $\alpha_{2}$ exchanged with}\nonumber\\\texttt{ an overall minus sign},
\end{eqnarray}
where the index $\alpha_{1}$ and $\alpha_{2}$ runs over particles states and $e$ is the single-particle energy. The core polarization parts are calculated
by keeping the intermediate states up to the 2$p$1$f$-shells. The single-particle matrix element reduced in both spin and isospin is written in terms of
the single-particle matrix element reduced in spin only~\cite{PM77}.
\begin{eqnarray}
\langle\alpha_{2}|||T_{\Lambda}|||\alpha_{1}\rangle=\sqrt{\frac{2T+1}{2}}\sum_{t_{z}}I_{T}(t_{z})\langle\alpha_{2}||T_{\Lambda}||\alpha_{1}\rangle
\end{eqnarray}
with
\begin{equation}
I_{T}(t_{z})=\left\{\begin{array}{ll}
1, & \quad\texttt{for}~~T=0, \\
(-1)^{1/2-t_{z}}, & \quad\texttt{for}~~T=1, \\
\end{array}\right \}
\end{equation}
where $t$$_{z}$=1/2 for protons and -1/2 for neutrons. The reduced single-particle matrix element of the Coulomb operator is given by~\cite{TJ66}
\begin{eqnarray}\
\langle\alpha_{2}||T_{J}||\alpha_{1}\rangle=\int^{\infty}_{0}\!dr\,r^{2}\,j_{J}(qr)
\langle\alpha_{2}||Y_{J}||\alpha_{1}\rangle\,R_{n_{1}\ell_{1}}\,R_{n_{2}\ell_{2}}
\end{eqnarray}
where $j_{J}(qr)$ is the spherical Bessel function and $R_{n\ell}(r)$ is the single-particle wave function. Electron scattering form factor involving
angular momentum $J$ and momentum transfer $q$, between the initial and final nuclear shell model states of spin $J_{i,f}$ and isospin $T_{i,f}$
is~\cite{TW84}
\begin{eqnarray}\
|F_{J}(q)|^{2}=\frac{4\pi}{Z^{2}(2J_{i}+1)}\left|\sum_{T=0,1}\left(%
\begin{array}{ccc}
T_{f} & T & T_{i} \\
-T_{z} & 0 & T_{z} \\
\end{array}%
\right)\right|^{2}\nonumber\\
\times\left|\langle\alpha_{2}|||T_{\Lambda}|||\alpha_{1}\rangle\right|^{2} |F_{c.m}(q)|^{2} \ |F_{f.s}(q)|^{2}
\end{eqnarray}
where $T_{z}$ is the projection along the z-axis of the initial and final isospin states and is given by $T_{z}=(Z-N)/2$. The nucleon finite-size (f.s)
form factor is $F_{f.s}(q) =\texttt{exp}(-0.43q^{2}/4)$ and $F_{c.m}(q) = \texttt{exp}(q^{2}b^{2}/4A)$ is the correction for the lack of translational
invariance in the shell model. $A$ is the mass number, and $b$ is the harmonic oscillator size parameter. The single-particle energies are calculated
according to~\cite{PM77}
\begin{eqnarray}
e_{nlj} &=& (2n+l-1/2)\hbar\omega\nonumber \\
&+&\left\{\begin{array}{ll}
-\frac{1}{2}(l+1)\langle~f(r)\rangle_{nl}, & \quad\texttt{for}~j=l-1/2, \\
\frac{1}{2}l\langle~f(r)\rangle_{nl}, & \quad\texttt{for}~j=l+1/2, \\
\end{array}\right \}
\end{eqnarray}
with $\langle~f(r)\rangle_{nl}\approx-20A^{-2/3}$ and $\hbar\omega=45A^{-1/3}-25A^{-2/3}$. The electric transition strength is given by~\cite{PM77}
\begin{equation}\
B(CJ,k)=\frac{Z^{2}}{4\pi}\left[\frac{(2J+1)!!}{k^{J}}\right]^{2}\,F^{2}_{J}(k)
\end{equation}
where $k=E_{x}/\hbar\,c.$
\section{Results and Discussion}
The core polarization effects are calculated with the MSDI as a residual interaction. The parameters of the MSDI are denoted by $A_{T}$ , $B$ and
C~\cite{PM77}, where $T$ indicates the isospin (0,1). These parameters are taken to be $A_{0}=A_{1}=B=25/A$ and $C=0$, where $A$ is the mass number. In all
of the following diagrams (see Fig. 1), the dashed lines give the results obtained using the 1$p$-shell wave functions (1p) of Cohen-Kurath interaction
[CK-TBME]~\cite{CK65}. The results of the core polarization (CP) effects are shown by the dashed-dotted lines. The results including core polarization
(1p+CP) are shown by the solid lines. The $B(C2\uparrow,q)$ values as a function of momentum transfer $q$ achieved by removing from the form factors most
of the dependence on the momentum transfer, according to the transformation given in Ref.~\cite{RA83}. The $B(C2\uparrow)$ values are given at the photon
point defined at $q = k = E_{x}/\hbar\,c$, and are displayed in Table~\ref{tab1}. The size parameter $b$ is taken to be 1.71 fm~\cite{EV87} to get the
single-particle wave functions of the harmonic-oscillator potential.

The calculations for the $C0$ and $C2$ isoscalar transition from the ground state ($J^{\pi}_{i} = 3^{+}, T = 0)$ to the ground state ($J^{\pi}_{f} =
3^{+}, T = 0)$ at $E_{x} = 0.0$ MeV are shown in Fig. 1. The multipole decomposition is displayed as indicated by $C0$ and $C2$. The total form factor is
shown by the solid curve, where the data are well described in all the momentum transfer regions up to $q\leq\,2.58 fm^{-1}$. The core-polarization
effects enhance the $C2$ form factor appreciably by a factor around 2 over the 1$p$-shell calculation. This enhancement brings the total form factor
(solid curve) very close to the experimental data. Similar results are obtained in Ref.~\cite{CD95}.

Fig.2 displays the calculation of the $C2$ form factor to the ($J^{\pi}_{f} = 1^{+}, T = 0)$ at $E_{x} = 0.718$ MeV. The 1$p$-shell model calculation
underestimate the experiment and the inclusion of the core polarization enhances the calculations and brings the form factor to the experimental values in
all momentum transfer regions.The result of the 1p-shell model calculations predicts the $B(C2\uparrow)$ value to be 0.889 $e^{2}fm^{4}$ in comparison
with the measured value $1.7\pm0.3\,e^{2}fm^{4}$~\cite{CD95}. Inclusion of CP effect predicts the value to be $1.77 e^{2}fm^{4}$, which is very close to
the measured value and the previous theoretical work of Refs.\cite{RA01,CD95} as shown in Table~\ref{tab1}.

The $C2$ form factor for the ($J^{\pi}_{f}T_{f}=2^{+} 0)$ at $E_{x}$=3.587 MeV is shown in Fig.3, the 1$p$-shell model calculations describes the
experimental data very well up to momentum transfer $q\leq\,2.0 fm^{-1}$ and start to deviate from the experiment. The inclusion of the core-polarization
effect overestimate the measured form factors up to $q\sim2.0 fm^{-1}$ and comes to the measured form factors at $q\sim(2-3) fm^{-1}$. The calculated
$B(C2\uparrow)$ value is found to be equal to 0.568 $e^{2}fm^{4}$ (without CP) and 1.55 $e^{2}fm^{4}$ (with CP) in comparison with the measured value
$0.6\pm0.1~e^{2}fm^{4}$~\cite{CD95} as displayed in Table~\ref{tab1}.

Fig.4\;shows the comparison of the calculated longitudinal $C2$ form factors from the ground state ($J^{\pi}_{i} = 3^{+}_{1}, T = 0)$ to the excited state
($J^{\pi}_{f} = 3^{+}_{2}, T = 0)$ at $E_{x} = 4.774$ MeV the 1$p$-shell model calculations reproduce the low-$q$ values up to $q\leq 1.0 fm^{-1}$ and
start to deviate severely and the inclusion of the CP effects make the calculations more worse and bring it higher than (1p) calculation. Our calculation
are consistent with that of Ref.~\cite{CD95} and in order to fit the measured form factor they use oscillator wave function with size parameter
$b=1.5\:fm$ and shows the $q$ dependance of the 8.66 MeV $C0$ form factor in $^{13}$C, normalized to fit the 4.774 MeV $^{10}$B data. The comparison of
the calculated $B(C2\uparrow)$ found to be 0.56 $e^{2}fm^{4}$ with (1p) and 1.66 $e^{2}fm^{4}$ with (1p+CP) in comparison with the measured value
$<0.04~e^{2}fm^{4}$~\cite{CD95}.

The longitudinal $C2$ form factor for the transition ($J^{\pi}=4^{+},T=0)$ state at $E_{x}=6.025\,MeV$, the inclusion of the core polarization effect
describes the measured form factor in all momentum transfer regions. The calculation of $B(C2\uparrow)$ with (1p) is found to be 5.79 $e^{2}fm^{4}$,
while with (1p+CP) is 11.67 $e^{2}fm^{4}$ in comparison with the measured value $17.4\pm0.7~e^{2}fm^{4}$~\cite{CD95} as shown in Fig.5 and
Table~\ref{tab1}. It is very clear that the 1p-shell model fails to describe the data in both the transition strength $(B(C2\uparrow) = 5.79 e^{2}
fm^{4})$ and the form factors. The inclusion of CP effects gives a remarkably good agreement with the experimental data in all regions of the momentum
transfers $q$ and enhances by a factor of 3 over the 1p-shell model results.
\begingroup
\begin{table}
\caption{\label{tab1} Theoretical values of the reduced transition probabilities $B(C2\uparrow,q)$ (in units of $e^{2} fm^{4})$ in comparison with
experimental values and other theoretical calculations for $^{10}$B.}
\begin{ruledtabular}
\begin{center}
\begin{tabular}{cccccccc}
$J^{\pi}_{f}$ & $T_{f}$ & $E_{x}$(MeV)& $b$ (fm) & 1$p$ & $1p$+CP & Other & Exp. \\
& & & \cite{EV87}& & &\cite{CD95} & \cite{CD95} \\
\hline\\
$3^{+}_{1}$ & 0 & 0.000 & 1.71 & & & & \\
$1^{+}$ & 0 & 0.718 & 1.71 & 0.889 & 1.77 & 1.62 & $1.7\pm0.3$\\
$2^{+}$ & 0 & 3.587 & 1.71 & 0.568 & 1.55 & 1.36 & $0.6\pm0.1$ \\
$3^{+}_{2}$ & 0 & 4.774 & 1.71 & 0.56 & 1.66 & 1.63 & $<0.04$ \\
$4^{+}$ & 0 & 6.025 & 1.71 & 5.79 & 11.67& 11.74& $17.4\pm0.7$ \\
\end{tabular}
\end{center}
\end{ruledtabular}
\end{table}
\endgroup
\begin{figure}
\centering
\includegraphics[width=0.4\textwidth]{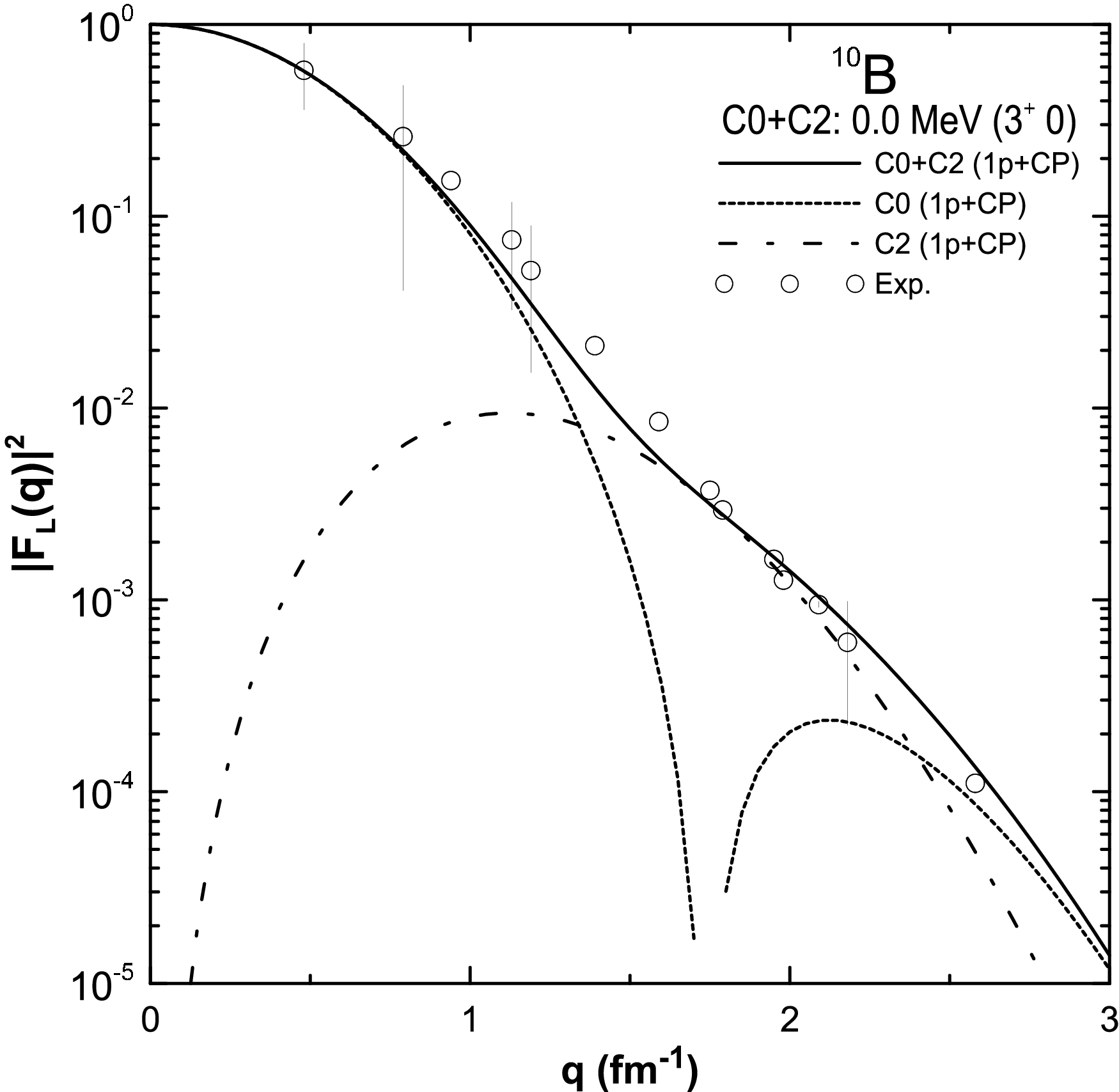}
\caption{The longitudinal $C0+C2$ form factor for the isoscalar $3^{+}_{g.s.}$ (0.0 MeV) transition in $^{10}$B compared with the experimental data taken
from Ref.~\cite{CD95}.}
\end{figure}
\begin{figure}
\centering
\includegraphics[width=0.4\textwidth]{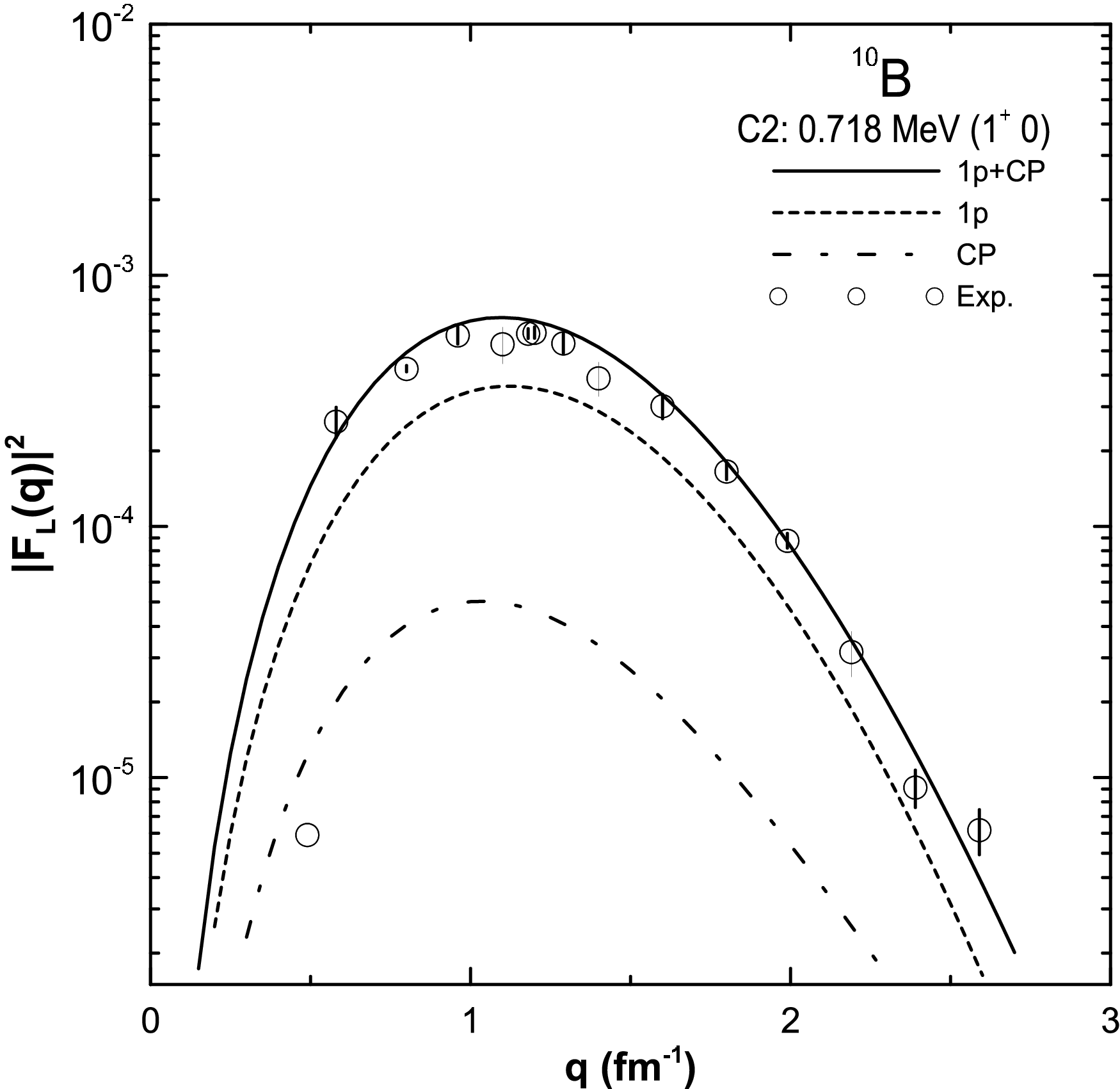}
\caption{The longitudinal $C2$ form factor for the isoscalar $1^{+}$ (0.718 MeV) transition in $^{10}$B compared with the experimental data taken from
Ref.~\cite{CD95}.}
\end{figure}
\begin{figure}
\centering
\includegraphics[width=0.4\textwidth]{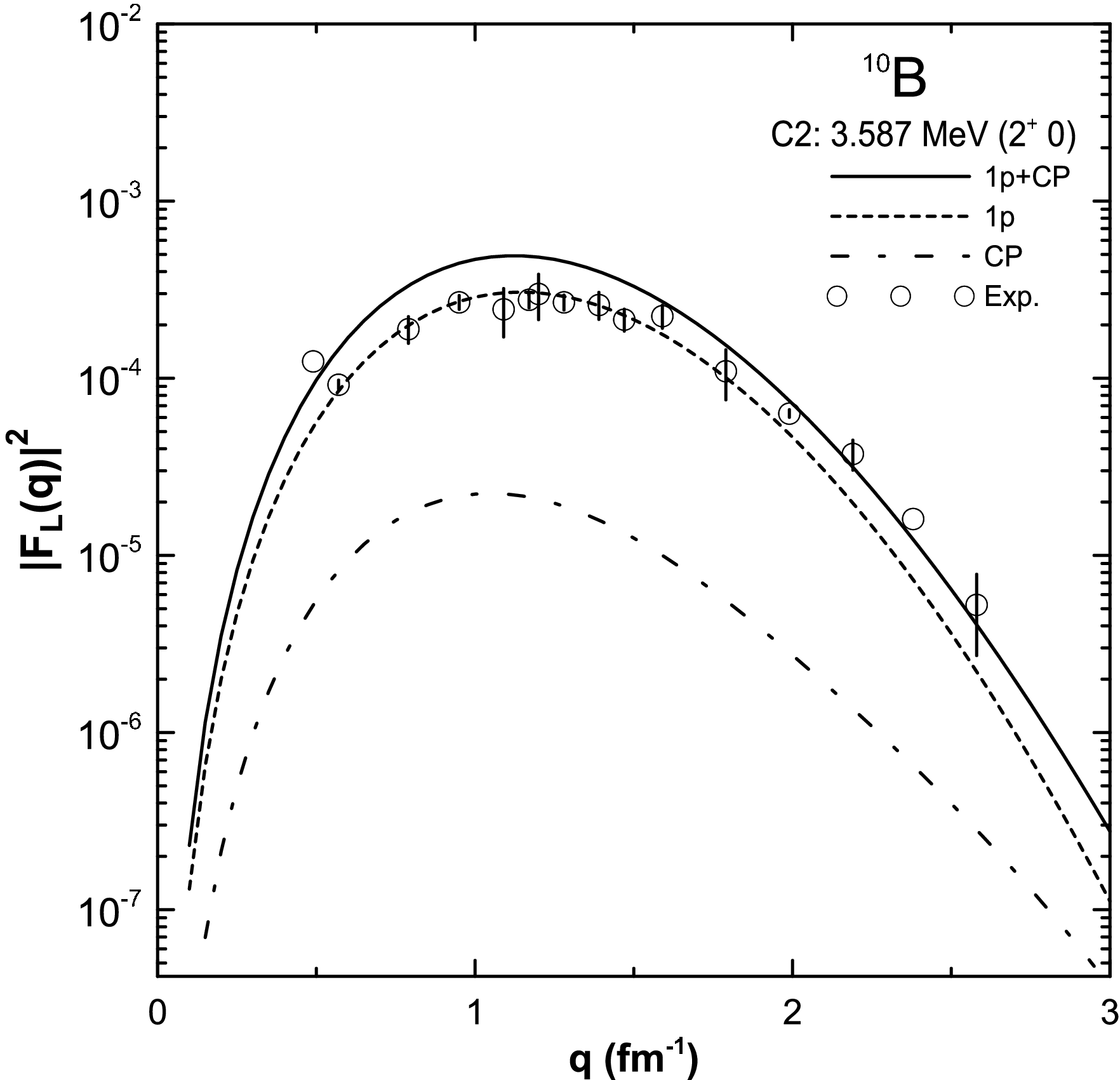}
\caption{The longitudinal $C2$ form factor for the isoscalar $2^{+}$ (3.587 MeV) transition in $^{10}$B compared with the experimental data taken from
Ref.~\cite{CD95}.}
\end{figure}
\begin{figure}
\centering
\includegraphics[width=0.4\textwidth]{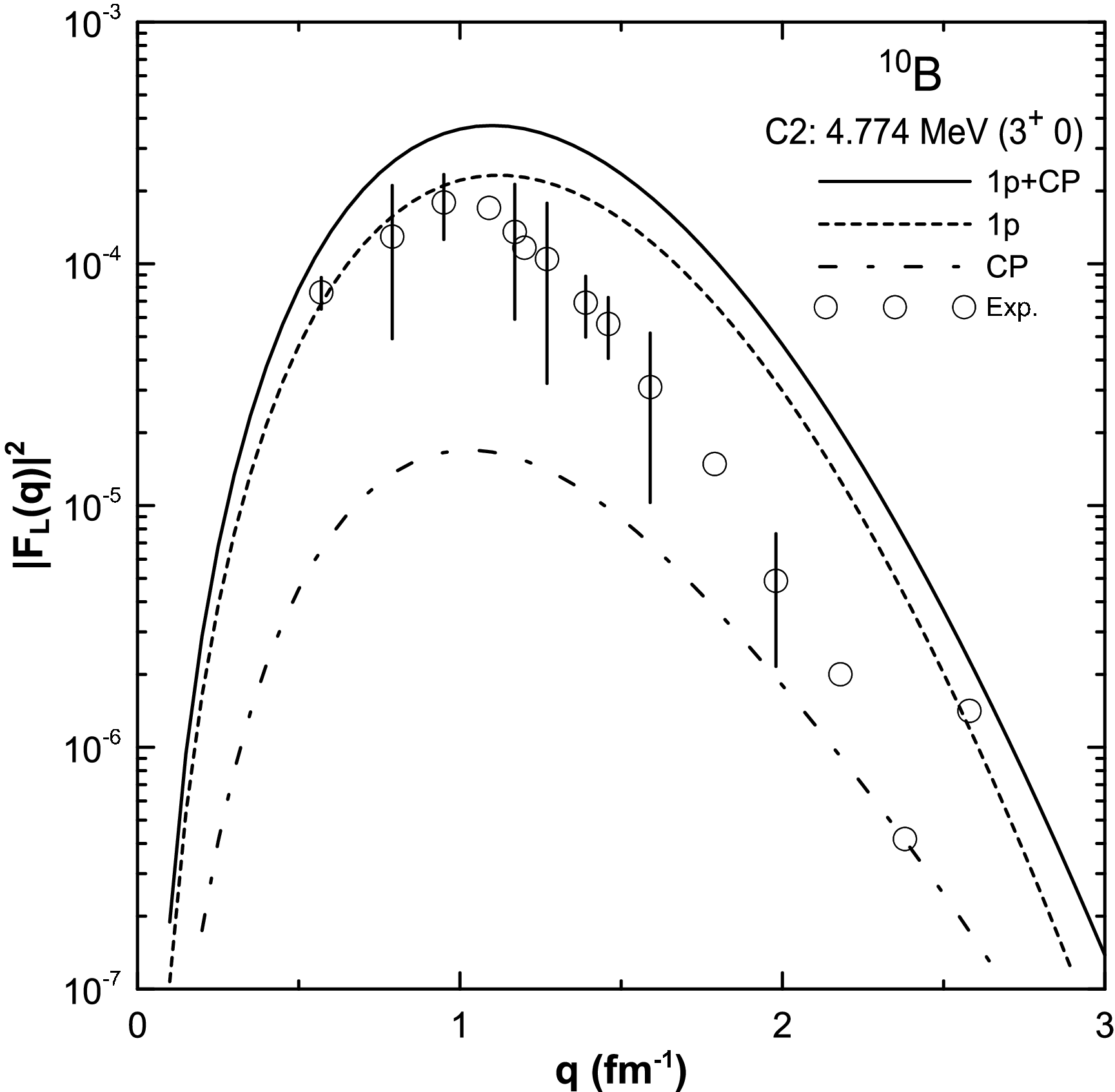}
\caption{The longitudinal $C2$ form factor for the isoscalar $3^{+}_{2}$ (4.774 MeV) transition in $^{10}$B compared with the experimental data taken from
Ref.~\cite{CD95}.}
\end{figure}
\begin{figure}
\centering
\includegraphics[width=0.4\textwidth]{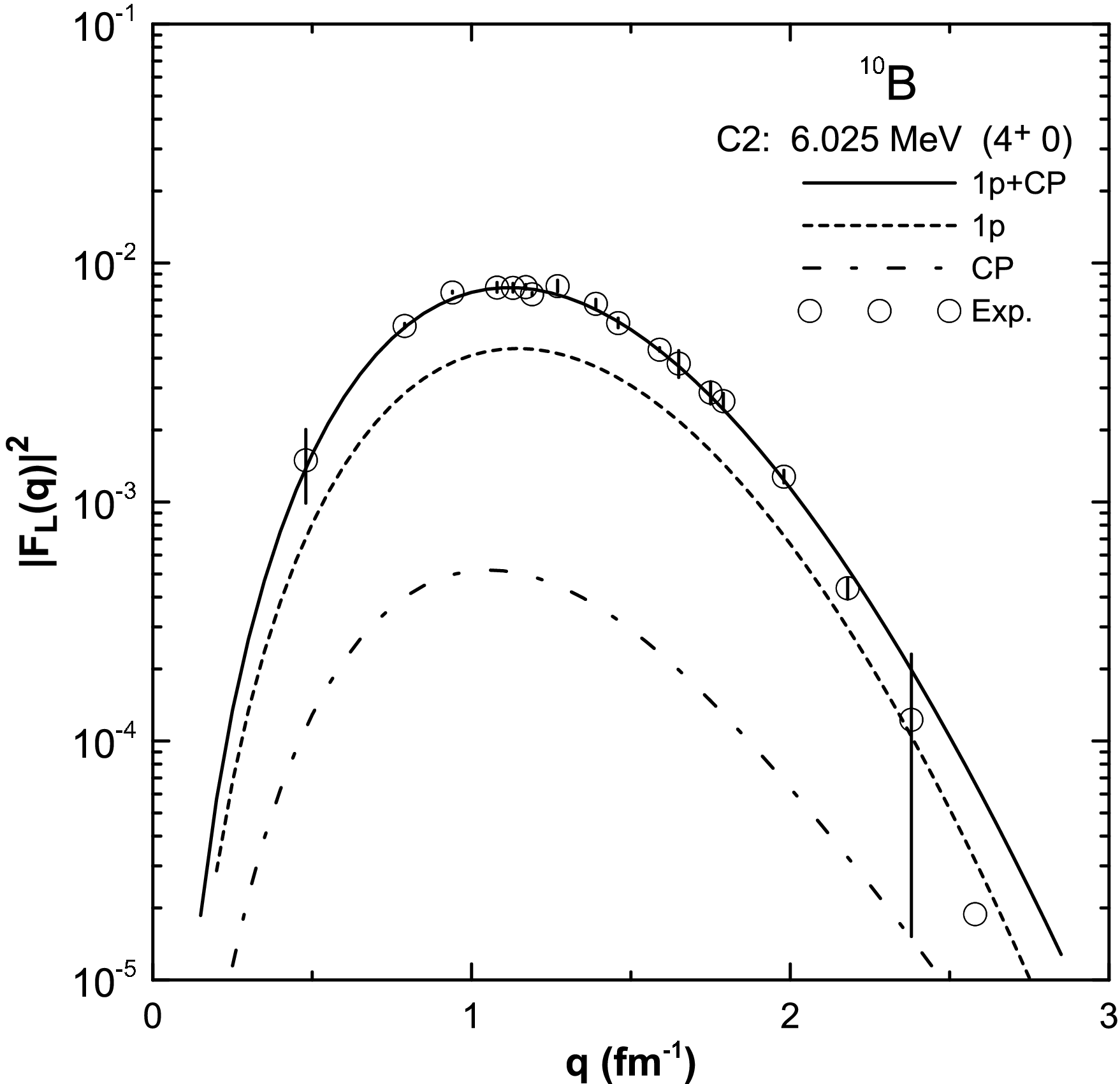}
\caption{The longitudinal $C2$ form factor for the isoscalar $4^{+}$ (3.587 MeV) transition in $^{10}$B compared with the experimental data taken from
Ref.~\cite{CD95}.}
\end{figure}
\newpage
\section{Conclusions}
The 1p-shell models, which can describe static properties and energy levels, are less successful for describing dynamic properties such as $C2$ transition
rates and electron scattering form factors. The average underestimation of the $B(C2\uparrow)$ value from the experiment is about a factor of 2. The
inclusion of higher-excited configurations by means of core polarization enhances the form factors and brings the theoretical results closer to the
experimental data. The average $B(C2\uparrow)$ value becomes about 90\% of the average experimental value when core polarization effects are included, for
the transitions considered in this work. All calculations presented in this work have been performed by employing MSDI as residual interaction. The use of
modern effective interaction may give a better description of the form factors.
\newpage

\end{document}